\documentstyle[12pt]{article}

\def\fnote#1#2 {\begingroup \def \thefootnote {#1}
\footnote{#2}\addtocounter{footnote}{-1}\endgroup}

\newcommand{\support} {This work is supported in part by U.S. Department of
Energy Contract No.  DE-FG02-91ER40685.}
\newcommand{\myname} {\vspace{0.5in}
                                  \begin{center}Zhu Yang\\
                                  \vspace{0.2in}
                                  Department of Physics and Astronomy\\
                                  University of Rochester\\
                                  Rochester, NY 14627\\
                                  bitnet:yang@uorhep\\
                                 \vspace{0.5in}
                                 ABSTRACT\\
                                  \vspace{0.2in}
                                \end{center}}

\newcommand{\pagenumber}{\pagestyle{plain}\setcounter{page}{1}}

\def\a{\alpha} 
\def\b{\beta} 
 
\def\e{\epsilon}

 \def\L{\Lambda}
\def\m{\mu} 
\def\n{\nu}

\def\s{\sigma} 
\def\t{\tau}

\def\raisenot{\raise .5mm\hbox{/}}
\newcommand{\notpa}{\hbox{{$\partial$}\kern-.54em\hbox{\raisenot}}}
\def\notp{\ \hbox{{$p$}\kern-.43em\hbox{/}}}
\def\notq{\ \hbox{{$q$}\kern-.47em\hbox{/}}}
\def\notk{\ \hbox{{$k$}\kern-.47em\hbox{/}}}
\def\notA{\ \hbox{{$A$}\kern-.47em\hbox{/}}}
\def\nota{\ \hbox{{$a$}\kern-.47em\hbox{/}}}
\def\notb{\ \hbox{{$b$}\kern-.47em\hbox{/}}}
\def\notD{\ \hbox{{$D$}\kern-.50em\hbox{/}}} %slash

\begin{document}
\baselineskip=24pt

\pagestyle{empty}

\begin{flushright}
UR-1288\\
ER-40685-737\\
\end{flushright}
\vspace{0.5in}
\begin{center}
{\Large  Asymptotic Freedom and Dirichlet String Theory}
\end{center}
\myname

Fixed angle  scattering at high energy in
a string theory  with boundaries satisfying Dirichlet conditions
(Dirichlet strings) in $D=4$ is shown
to have logarithmic dependence on energy, in addition to the power-like
behavior known before. High temperature free energy also depends
logarithmically on temperature.
Such a result could provide a
matching mechanism  between strings at long distance  and asymptotic
freedom at short distance, which is
necessary for the reformulation
of large-$N$ QCD as a string theory.

\newpage
\pagenumber

In order to reformulate the large-$N$ QCD \cite{th}
as a string theory, assuming
it is possible at all, we must find a string theory that contains
only massive excitations (for pure QCD) and shares the
characteristic features of asymptotic freedom, such as
partonic behavior and logarithmic violation of scaling at
high energy. The critical
string theories we are mostly familiar with
fail to meet theses two requirements: they contain massless gravitons and
massless Yang-Mill particles, and fixed angle scattering falls off
exponentially with total energy.
%It is not clear whether these two phenomena are intimately related.
Liouville theory \cite{p1} has been argued to be irrelevant for the problem
of QCD \cite{jp}, meanwhile the effective string picture advocated in
\cite{ps}, which without doubt is correct for long strings,
 has problems at short distances.
Rigid strings \cite{pk} at first sight seem to give the
right behavior at
high energy and high temperature, but they have the wrong sign
of free energy at high temperature as compared with QCD \cite{joe},
and furthermore there are problems with unitarity \cite{py}.

Another simple yet non-trivial modification of short distance behavior
of strings is to add world-sheet boundaries satisfying Dirichlet
boundary conditions -- each boundary is mapped to a single point
in space-time and the position is then integrated over.
Such a theory was originally proposed as an off-shell extension of
closed string amplitudes \cite{gr}. Meanwhile the point-like states
of the Dirichlet boundaries were proven to produce power-like
behavior of fixed angle scattering at high energy \cite{gr2}.
The point-like structure was then linked to partons.
Developments of the Dirichlet string theory in modern
language can be found in \cite{gr3}.
Recently Green \cite{gr4} points out that at high temperature the Dirichlet
boundaries produce a free energy similar to that of the
large-$N$ QCD found by Polchinski\cite{joe}. Namely, in QCD the free
energy of a long string per unit length goes as
\begin{equation}
\mu_{QCD}^{2}(\b)/L^{2} \sim -{\frac{2g^{2}(\b)N}{\pi^{2}\b^{4}}},
\end{equation}
while for the Dirichlet string,
\begin{equation}
\mu_{DS}^{2}(\b)/L^{2} \sim -{\frac{8\pi^{2} w_{D}
g_{open}^{2}(\b)}{\b^{4}}},
\end{equation}
where $w_{D}$ is related to Chan-Paton factor and  here is chosen to be
$N$. We should think $w_{D}g_{open}^{2}$ as the weight of inserting a
Dirichlet boundary. It is quite natural to study this string
theory further to determine whether it has anything to do with
large-$N$ QCD.

The QCD result  does not simply imply that the free energy is
some power of temperature $1/\b$. The gauge coupling constant $g^{2}(\b)$
in $D=4$ depends  logarithmically on the scale one is probing.
The asymptotic freedom \cite{dwp} is the most distinctive feature of QCD.
It is the purpose  of this paper to explore the
logarithmic renormalization in the Dirichlet string theory, which
was anticipated in Green's work \cite{gr3}.
This study is interesting because it could provide a mechanism to
make string theory ``unstringy" at short distance, thus makes contact
with asymptotically free large-$N$ gauge theories: it can
be a matching condition from a string theory and QCD, much as
matching two effective field theories at some common scale.
The particular model we are going to work on may of course not be the
correct one, but it illustrates the point.  Since this is a
first work on higher order corrections in the Dirichlet string theory,
we will learn  how short distance behavior
gets modified by loop effect.

The model we pick up is a bosonic critical string theory
compactified  on a torus  $T^{22}$ of radius $R$ in each direction,
 with insertions  of world sheet boundaries
satisfying Dirichlet conditions for 4 large dimensions and
usual Neumann conditions for the rest 22  dimensions.
The string coordinates are thus denoted by $X^{\mu}(\s,\t)$
($\m = 1, ..., 4$) and $X^{I}(\s,\t)$ ($I = 5, ... 26$).
At the boundary $\partial M$  of the world sheet $M$,  roughly
speaking,
$X^{\m}|_{\partial M} = x_{b}^{\m}, \partial_{n}X^{I}|_{\partial M} =0$,
here $\partial_{n}$ means normal derivative.
These boundaries are literally true instantons:
they live for infinitely short duration in time.

The Polyakov functional integral\cite{p1}   for
strings with Dirichlet boundary conditions was studied in
\cite{oa}. We will follow  the treatment of \cite{cmnp} for
boundary conditions.
Let us first look at the basic
physical picture of Dirichlet strings \cite{gr2,gr3}.
Amplitudes in ordinary closed string theory is formulated as sum
over closed
riemann surfaces with punctured holes representing positions of
various vertex operators. A closed string coupling constant
$g_{c}^{2}$ is added to distinguish contributions from different genus.
In the present case the world sheet consists of additional boundaries,
each of which is mapped  to a single and different  space time point.
Those points are then integrated over the whole space time.
It is important that the weight   ($w_{D}g_{open}^{2}$ in (2))
of adding one such boundary is
independent of $g_{c}^{2}$, while in ordinary open string theory
the two are related due to unitarity.
Thus we think about the Dirichlet string theory as a
theory of $closed$ strings
or ``glueballs". One can of course simultaneously
 have other boundaries
with Neumann conditions, which would be thought as adding ``mesons".

Assuming the Riemann surface we are considering has $B$ boundaries,
$E$ vertex operators, and genus $G$,
according to the general procedure of evaluating the Polyakov
path integral\cite{oa,p4,mn}, we have for
the string amplitude
\begin{eqnarray}
A_{E}(k_{1} ... k_{E}) &=&  C_{M} g_{c}^{2G+E} (w_{D}g^{2}_{open})^{B}
\prod_{b=1}^{B}\int d^{D}x_{b}  \int_{M}d^{n}\t  [det^{\prime}
P^{\dagger}P]^{1/2} [{\frac{det H(P^{\dagger})}{detH(P)}}]^{1/2}\,
\nonumber \\
&&\int DX e^{-S(x, \hat{g})}
\prod_{e=1}^{E} \int d^{2}\s_{e}V_{e}(\s_{e}, k_{e}).
\end{eqnarray}
The only difference from the usual string theory is the integration over
$x^{\m}_{b}$, apart from
a trivial change of boundary conditions.
We do not specify various symbols because they are either
self-evident or will be given in specific cases later.

Let us comment on the operator formulation, which gives
clearer physical picture \cite{gr3}. Due to duality or conformal
invariance, we can think of the
string amplitude as a diagram with closed string propagating (with
some initial or final boundary states) or with open string
in the intermediate channel.
The former interpretation uses the familiar closed string propagator
$(L_{0} + \bar{L}_{0} -2)^{-1}$, while the latter is more interesting.
$X(\s, \t)$ of an open string satisfying the above boundary condition is
given by
\begin{eqnarray}
X^{\m} &=& x_{1}^{\m} + (x_{2}^{\m}-x_{1}^{\m}){\frac{\s} {\pi}}
+ \sqrt{2\a^{\prime}} \sum_{n} {\frac{1}{n}}\a_{n}^{\m}
 \sin n\sigma e^{-in\t}\, ,
\nonumber \\
X^{I}  &=& x^{I} + {\frac{2\a^{\prime}m^{I}}{R}} \t +
\sqrt{2\a^{\prime}} \sum_{n} {\frac{1}{n}}\a_{n}^{I}
 \sin n\sigma e^{-in\t}.
\end{eqnarray}
$L_{0}$ is then
\begin{equation}
L_{0} = {\frac{1}{4\a^{\prime}\pi^{2}}} [(x_{2} - x_{1})^{2}
+ (2\a^{\prime} m/R)^{2}] + \sum_{n=1}^{\infty}\a_{-n}\cdot
\a_{n}.
\end{equation}
The propagator is $(L_{0} - 1 )^{-1}$.
The first term in (5) is most notable:
it replaces the usual $p^{2}$. The reader should consult \cite{gr3}
for more details.

The lowest order contribution of the instantons is of course
one boundary insertion. Various aspects  have been discussed thoroughly
by Green \cite{gr3,gr4}.
The world sheet in question has the topology of a disk.
One can show that two tachyon scattering
amplitude at fixed angle and high energy is proportional to
$(stu)^{1/2}$, where $s, \,t$ and $u$ are Mandelstam variables.
The growth of amplitudes at high energy is related to the existence of
tachyons in the dual theory, and is expected to change when
the correct vacuum is found. The high temperature free
energy is already given in (2).

Our aim is to consider two instanton contribution.
This is similar but physically distinct from the semi-off-shell amplitudes
considered in \cite{cmnp1}.
The world sheet in
a cylinder or annulus.
The two boundaries are mapped to $x_{1}^{\m}$
and $x_{2}^{\m}$, respectively.
We are interested in the case when the
two instantons are close in $spacetime$, and they can be
seen as  a single instanton with renormalized strength. Conceptually
this looks very much like how renormalization works in
field theory.  Since the bosonic string theory
has various divergences, one
must be careful about their interpretations. We are not worried
about divergences caused by the closed string  states, since
they are related to the infrared behavior of the theory and can
be cured.  Rather the interesting divergences occur in the
open string channel. In (5) there is a singularity outside the
light cone, which is the reminiscence of open string tachyon in the
dual theory. We treat this as we do with the closed string tachyon,
that is, by analytical continuation.
More interesting divergence occurs on the light cone.
Formally if we take $\a^{\prime}$ to infinity, only light cone
singularity remains \cite{gr3}. It is interesting that this is also the
limit we would like to consider in high energy scattering \cite{gm}.
%The real justification should ultimately come from the
%renormalization group.
%at least from the fact that only logarithmic
%divergence at short distances  implies
%nontrivial dependence on energy, which can not be eliminated by
%local counterterms.

We first give a heuristic argument for the logarithmic renormalization
in the Dirichlet theory, based on analyticity in ``position space"
studied in \cite{gr3}. We visualize a cylinder diagram as sewing
of several open string propagators (5) with some vertices that
couple the open string to closed string states,  which
are tachyons here. Now look at a particular channel where the
world sheet consists of a long tube along which two  open strings
propagate, and closed
string states at each side. The propagators of the two open strings
are the same as (5). So we have the following term
\begin{equation}
\int d^{4}\Delta x {\frac{1}{(L_{0}-1)^{2}}},
\end{equation}
where $\Delta x^{\m} = x^{\m}_{2} - x^{\m}_{1}$.
Take the first excited state in $L_{0}$, we get logarithmic divergence.
This picture also prompts us to look at suitable corners of moduli
space in more refined treatment.

We  begin a detailed discussion
of fixed angle two tachyon scattering amplitude at high energy on
an  annulus or cylinder. In (3),
we need to know some determinants and the propagator
on the world sheet.
The determinants are neatly worked out in both
\cite{cmnp} in the Polyakov approach and in \cite{gr3} in operator
formalism.  The cylinder has one Teichmuller parameter $l$ and
we choose our background metric to be
$\hat{g}_{ab} d\s^{a}d\s^{b} = (d\s^{1})^{2}+  l^{2} (d\s^{2})^{2}$.
Both $\s^{1}$ and $\s^{2}$ take range $[0, 1]$.
The boundaries are at $\s_{2}=0, 1$.
The vertex operators are
\begin{equation}
\prod_{i=1}^{E} \int d^{2}\s_{i} \sqrt{\hat{g}(\s_{i})} \exp
ik_{i\m}X^{\m}(\s_{i}),
\end{equation}
where $k_{i}^{2}=8\pi T = 4/\a^{\prime}$. The integration variables
are $x_{1}, x_{2}$ and $l$.
The integration measure in (3) is, including contributions from
matter sector,
\begin{equation}
{\frac{1}{(\a^{\prime}l)^{4}}}
F_{1}^{22}(a, q) f(q^{2})^{-24}
\exp[-(x_{2}-x_{1})^{2}/(4\pi\a^{\prime}l) + 4\pi l +
i\s^{2}_{i}p_{i}\cdot
(x_{2}-x_{1}) + i p_{i}\cdot x_{1}],
\end{equation}
where we have used the standard notations $q = \exp(-2\pi l)$,
$a^{2} = \a^{\prime} / R^{2}$, and \cite{gsw}
\begin{equation}
F_{1}(a, q) = \theta_{3}(0|{\frac{\ln q}{2\pi i a^{2}}}) =
\sum_{-\infty}^{\infty} q^{n^{2}/2a^{2}}, \,\,\,\,\,\,
f(q^{2}) = \prod_{n=1}^{\infty}(1-q^{2n}).
\end{equation}
$F_{1}^{22}(a, q)$ arises from sum over the zero modes or momentum states
of compactified dimensions.
$e^{4\pi l}f^{-24}(q^{2})$ comes from oscillator summation
of both matter and ghost.
Notice the Dirichlet boundary condition prevents usual zero modes
in uncompactified dimensions. This shows up as the factor $1/l^{D}$.
And finally the exponential of $x$ and $p$ terms is the contribution
of classical action.

We need to know the propagator on the cylinder. Since the tachyons
involve $X^{\m}$ only, we just give the expression for
Dirichlet condition $X^{\m}(\s^{1},0)=X^{\m}(\s^{2}, 1) = 0$. Note
we have already extracted the zero mode contribution in (8).
The propagator is easily found by the method of images to be \cite{cmnp1}
\begin{equation}
\langle X(\s_{i})X(\s_{j})\rangle
= -{\frac{ l\s^{2}_{i}\s^{2}_{j} }{T} }
-{\frac{1}{2\pi T}} \ln | {\frac{\theta_{1}(\n_{ij}|2li)}
{\theta_{1}(\bar{\n}_{ij}|2li)}}|,
\end{equation}
where $\n_{ij} = \s_{i}^{1}-\s_{j}^{1} + il (\s_{i}^{2}-\s_{j}^{2})$
and $\bar{\n}_{ij} = \s_{i}^{1}-\s_{j}^{1} + il (\s_{i}^{2}+\s_{j}^{2})$
are image points, and $\theta_{1}(z|\t)$ is the Jacobi $\theta$-function
(see, e.g.,\cite{gsw}).
When $i = j$, we make one subtraction \cite{p4}, and get
\begin{equation}
\langle X(\s_{i})X(\s_{i})\rangle = -{\frac{ l(\s^{2}_{i})^{2}}{T}}
-{\frac{1}{2\pi T}} \ln | {\frac{\theta_{1}^{\prime}(0|2li)}
{\theta_{1}(2l \s_{i}^{2} i|2li)}}|.
\end{equation}

We  perform integration over $x_{1}^{\m}$ first, which imposes
momentum conservation. We use the Mandelstam variables
$s=-(k_{1}+k_{2})^{2}, \, t=-(k_{1}+k_{3})^{2}$, and $u=-(k_{1}+k_{4})^{2}$.
Put (7)-(12) in to (3) and use the on-shell condition
we arrive at the expression
\begin{eqnarray}
A_{4} &=& C g_{c}^{4} (w_{D}g_{open}^{2})^{4}
\int d^{4}\Delta x \int_{0}^{\infty}dl \prod_{i=1}^{4}(l \int d^{2}\s_{i})
{\frac{1}{(\a^{\prime}l)^{2}}
F_{1}^{22}(a, q) f(q^{2})^{-24}}\, \nonumber \\
&&\exp[-(\Delta x)^{2}/(4\pi\a^{\prime}l) + 4\pi l+
i\s^{2}_{i}p_{i}\cdot
\Delta x ] \prod_{j=1}^{4}\lbrace \exp(2\pi l \s_{j}^{2})
| {\frac{\theta_{1}^{\prime}(0|2li)}
{\theta_{1}(2l \s_{j}^{2}|2li)}}|\rbrace^{2} \, \nonumber \\
&& \lbrace{\frac{\chi_{12}\chi_{34}}{\chi_{14}
\chi_{23}}}\rbrace^{-s/4\pi T}
\lbrace{\frac {\chi_{13}\chi_{24}}
{\chi_{14}\chi_{23}} }\rbrace^{-t/4\pi T}
\lbrace{\frac{\chi_{12}\chi_{13}\chi_{24}\chi_{34}}
{\chi_{14}\chi_{23}}}\rbrace^{-4},
\end{eqnarray}
where
\begin{equation}
\chi_{ij} = \exp \lbrace +2\pi l\s^{2}_{i}\s^{2}_{j}\rbrace
| {\frac{\theta_{1}(\n_{ij}|2li)}
{\theta_{1}(\bar{\n}_{ij}|2li)}}|.
\end{equation}
(12) coincides with \cite{cmnp1}, apart from the $\Delta x$ integration
and compactification.
We can of course integrate over $\Delta x^{\m}$ in (12), since
it is a Gaussian. That will
make closed string poles clear.
Indeed the way the amplitude is written is suitable for discussing
closed string intermediate states.
Since we are interested in possible
divergence in $\Delta x^{\m}$ integration, we will perform it at
last step. It is convenient to make Jacobi transformation now,
which is equivalent to a conformal transformation of the world
sheet metric by a factor $1/l^{2}$. (12) reads,
\begin{eqnarray}
A_{4} &=& C g_{c}^{4} (w_{D}g_{open}^{2})^{4}
\int d^{4}\Delta x \int_{0}^{\infty}
{\frac {dl^{\prime}}{l^{\prime}}} \prod_{i=1}^{4}(l^{\prime}
\int d^{2}\s_{i})
\theta_{3}^{22}(0|2\pi a^{2} l^{\prime}i)
f(e^{-2\pi l^{\prime}})^{-24}\, \nonumber \\
&& \exp[-(\Delta x)^{2} l^{\prime}/(2\pi\a^{\prime}) + 2\pi l^{\prime}+
i\s^{2}_{i}p_{i}\cdot
\Delta x ] \prod_{j=1}^{4}\lbrace
| {\frac{\theta_{1}^{\prime}(0|l^{\prime}i)}
{\theta_{1}( \s_{j}^{2}|l^{\prime}i)}}|\rbrace^{2} \, \nonumber \\
&& \lbrace{\frac{\chi_{12}\chi_{34}}{\chi_{14}
\chi_{23}}}\rbrace^{-s/4\pi T}
\lbrace{\frac {\chi_{13}\chi_{24}}
{\chi_{14}\chi_{23}} }\rbrace^{-t/4\pi T}
\lbrace{\frac{\chi_{12}\chi_{13}\chi_{24}\chi_{34}}
{\chi_{14}\chi_{23}}}\rbrace^{-4},
\end{eqnarray}
where  we have used $l^{\prime} = 1/2l$ and $\chi_{ij}$ changes to
\begin{equation}
\chi_{ij} =
| {\frac{\theta_{1}(-\n_{ij}l^{\prime}i|l^{\prime}i)}
{\theta_{1}(-\bar{\n}_{ij}l^{\prime} i|l^{\prime}i)}}|.
\end{equation}

It is clear that when $s, t$ are both large, the last
factor in (12) can be ignored, and the integral is dominated
by  saddle points \cite{gm}. In our case, they occur
when all $\s_{i}^{2}$'s approach
the two boundaries of the cylinder, so that $\chi_{ij}$ approaches
1. If we take one of the vertex to the boundary,
we find a $1/\e$ divergence, which is related to the singularity
outside the light cone \cite{gr3}. This divergence is
removed by redefining our vertex operators \cite{sw}.
In order to match the intuitive discussion before,
we will also take the limit $l \rightarrow 0 (l^{\prime}
\rightarrow \infty)$.
It is proven convenient to limit invariant distance between two
vertex operators or between one vertex operator and one boundary
to be larger than $\e$ \cite{sw}.

We take the limit $ \s^{2}_{i} \rightarrow 0$.
We simply Taylor expand various $\theta$-functions.
Use the fact that $\theta^{\prime}_{1}/\theta_{1} \sim i\pi$ and
$\theta^{\prime\prime}_{1}/\theta_{1} \sim - \pi^{2}$ for the range of
variables are are interested in, the $s$ and $t$ power terms simplify to
$\exp ( \sum a_{ij} \s^{2}_{i}\s^{2}_{j}s/4\pi T)$ and a similar
expressions for the $t$ term. If we take $\s^{2}\rightarrow 0$ faster than the
other three, and change variables to
\begin{equation}
z_{1} = \s^{2}_{3}\s^{2}_{4}, \,\,
z_{2} = \s^{2}_{2}\s^{2}_{4}, \,\,
z_{3} = \s^{2}_{2}\s^{2}_{3}, \,\,
\end{equation}
(14) becomes
\begin{equation}
A_{4} \sim ... \, \int {\frac {d\s^{2}_{1}} {(\s^{2}_{1})^{2} } }
\int {\frac {dz_{1}dz_{2}dz_{3}} {(z_{1}z_{2}z_{3})^{3/2} } }
\exp (z_{1}s/4\pi T + z_{2}t /4\pi T + z_{3}u/4\pi T)
\sim (stu)^{1/2}.
\end{equation}
The above integral is divergent, which is cut off by $\e$ introduced
before. One can add conterterms living on the boundaries to
to cancel the divergence
\cite{gr3,cmnp1}.

Now we do the remaining integrals.
When $\s_{i}^{1}$ are far apart, our treatment of $\theta$ functions
is correct. After completing the integral we get divergence in powers
of a space-time cutoff $|\Delta x | > 1/\L$.
When, say, $\s_{1}^{1}$ and $\s^{1}_{2}$, and similarly
$\s_{3}^{1}$ and $\s_{4}^{1}$ get close to the order of $1/l^{\prime}$,
respectively, there is a change. In this case it is convenient to use
$ \eta_{12}/l^{\prime}= \s_{1}^{1} - \s^{1}_{2}$ and
$ \eta_{34}/l^{\prime}= \s_{3}^{1} - \s^{1}_{4}$.
(12) is proportional to
\begin{equation}
\int d^{4}\Delta x \int_{0}^{\infty} {\frac {dl^{\prime}}{l^{\prime}}}
l^{\prime 2} \exp -[l^{\prime}(\Delta x)^{2}
/ 2\pi \a^{\prime} ],
\end{equation}
with again some $\e$ dependence.
Strictly speaking the above equation should include sum of terms
with $\exp (2\pi nl^{\prime})$ and
$\exp (2\pi a^{2} nl^{\prime})$, but they don't give rise
logarithmic divergence and as mentioned before we ignore them.
The above integral becomes
$\ln (\L^{2}/s)$.
The $s$ factor is natural on dimensional grounds.
It is not necessary for all four vertex operators to touch the
same boundary. For example, if $\s_{1}^{2}, \s_{2}^{1} \rightarrow
 1$ while the other two go to zero, there will be a factor
$\exp i(p_{1}+p_{2})\cdot \Delta x$ in (15) and (16), which
is also logarithmic divergent.

Let's write down the logarithmically
divergent contribution to the amplitude (12)
\begin{equation}
A_{4D} \sim C_{2} g^{4}_{c} (w_{D}g_{open}^{2})^{2} \ln (\L^{2}/s)
(stu)^{1/2}.
\end{equation}
If we renormalize the Dirichlet coupling constant
as
\begin{equation}
w_{D}g_{open}^{2}(\m)= w_{D}g_{open}^{2}(\L) + {\frac{C_{2}}{C_{1}}}
(w_{D}g_{open}^{2})^{2} \ln (\L^{2}/\m^{2}),
\end{equation}
where $C_{1}$ as a factor from the disk diagram,
we can write down the renormalization group equation
\begin{equation}
{\frac{d (w_{D}g_{open}^{2})}{d\ln \m}} = -{\frac{2C_{2}}{C_{1}}}
(w_{D}g_{open}^{2})^{2},
\end{equation}
which is characteristic of  asymptotically free
theories (assuming the positivity of $C_{1}$ and $C_{2}$).
This is the main result of this paper.

We  may also compute the
high temperature partition function by relating it to the mass
squared of winding tachyon states. This is actually simpler
than the one performed above, since the diagram involves only two
vertex operators.
We find the similar logarithmic divergence as (21), and we
conclude the winding tachyon mass  (2) contains logarithmic
dependence on $\b$ as well.

Some comments are in order.  First of all, during our investigation
we encountered power-like divergences at short
distance in spacetime. In absence of an off-shell
formulation, it is hard to deal with them. If the standard lore
of renormalization applies, we would conclude they are not interesting
because they can be cancelled by local counterterms.
Logarithmic divergence is entirely different matter, of course. It is
this that makes our result interesting. One interpretation of
the power-like divergences would be that the Dirichlet string theory
is too singular to be a consistent theory
at short distance. At best it is an effective
theory, as the pion phenomenological Lagrangian \cite{sw1}.
The logarithmic
dependence we found would serve as the matching condition between
long distance string theory and short distance QCD.
It is not ruled out, however, that the theory could
be fully consistent once its true vacuum is found, or some
other models, with for example more world sheet symmetries, are
considered.
We would like to see a more systematic treatment of
other divergences in this
string theory. To cancel these divergences we need to go beyond
Weyl invariant theory in general \cite{sw,cmnp1}.
So an off-shell formulation
is also needed here. Eventually this
will help us to understand mass correction, gauge invariance and
other issues.

In conclusion, the logarithmic divergence of Dirichlet string
theory could be an important indication that a string formulation
of large-$N$ QCD is possible. Certainly this deserves further study,
and hopefully a new consistent string theory will emerge.

I would like to thank J. Polchinski for suggestions and encouragement.
Conversation with D. Minic is also appreciated.
\support

\end{document}